\documentclass[aps,nofootinbib,twocolumn]{revtex4} 
\usepackage{graphicx}
\input epsf
\newcommand{\sfig}[2]{
\includegraphics[width=#2]{#1}
		}
\newcommand{\Sfig}[2]{
	\begin{figure}[thbp]
	\sfig{#1.eps}{0.9\columnwidth}
	\caption{{\small #2}}
	\label{fig:#1}
	\end{figure}
}
\newcommand{\Rf}[1]{\ref{fig:#1}}
\newcommand{\rf}[1]{\ref{fig:#1}}

\def\cmm2{{\,\rm cm^{-2}}}
\def\cm2{{\,{\rm cm}^2}}
\def\cmm3{{\,{\rm cm}^{-3}}}
\def\gcmm3{{\,{\rm g\,cm^{-3}}}}

\def\fun#1#2{\lower3.6pt\vbox{\baselineskip0pt\lineskip.9pt
  \ialign{$\mathsurround=0pt#1\hfil##\hfil$\crcr#2\crcr\sim\crcr}}}

\def\be{\begin{equation}}
\def\ee{\end{equation}}
\def\bea{\begin{eqnarray}}
\def\eea{\end{eqnarray}}

\newcommand{\ec}[1]{equation~[\ref{eq:#1}]}

\newcommand{\eql}[1]{\label{eq:#1}}

\begin{document}

\title{Cluster Masses Accounting for Structure along the Line of Sight} 

\author{Scott Dodelson$^{1,2}$}

\affiliation{$^1$NASA/Fermilab Astrophysics Center
Fermi National Accelerator Laboratory, Batavia, IL~~60510-0500}
\affiliation{$^2$Department of Astronomy \& Astrophysics, The University of Chicago, 
Chicago, IL~~60637-1433}

\date{\today}
\begin{abstract}
Weak gravitational lensing of background galaxies by foreground clusters offers an
excellent opportunity to measure cluster masses directly without using gas as a
probe. One source of noise which seems difficult to avoid is large scale
structure along the line of sight.
Here I show that, by using standard map-making techniques, one can minimize
the deleterious effects of this noise. The resulting uncertainties on cluster masses
are significantly smaller than when large scale structure is not properly accounted for,
although still larger than if it was absent altogether.
\end{abstract}
\maketitle

{\parindent0pt {\it Introduction.}} Clusters of galaxies are powerful
cosmological probes~\cite{cluster}. In particular, the number density of clusters as a function
of mass and redshift depends sensitively on the cosmological growth function which in turn
depends on the matter density and properties of the dark energy~\cite{wang}. Indeed, it is conceivable
that we will learn as much about dark energy from clusters as from supernovae~\cite{haiman}.

While redshifts of clusters are relatively easy to obtain, cluster masses are much
harder to pin down. A cluster mass can be estimated from optical richness or from 
its temperature, measured either in the X-ray
or with radio observations of the Sunyaev-Zel'dovich distortion of the cosmic microwave
background. However, the scatter around any of these indicators is large~\cite{scatter} and depends on complicated
physics, such as radiative transfer, star and galaxy formation, cooling, accretion, and feedback
mechanisms. In principle, gravitational lensing~\cite{bartel} allows for a more direct mass determination, 
since the distortions
of background galaxies are sensitive only to the mass along the line of sight, not to the
gas. In practice, there are many hurdles to overcome, most of which involve the observations
themselves.

There is one systematic that affects lensing determinations of clusters that cannot
be cured with better instruments or seeing. Large scale structure along the line of sight
can be mistaken as being part of the cluster~\cite{los}. As Hoekstra~\cite{hoekstra} has shown, the
uncertainties caused by this noise can significantly impair our ability to estimate
cluster masses. Since cluster abundances depend sensitively on their masses, this uncertainty
clouds the possiblity of learning about dark energy from clusters.

Here I show that one can partially offset the deleterious effects of
large scale structure by accounting for it in cluster mass estimates. 
To demonstrate how this works, I will focus on a single cluster
at redshift $0.3$ with a Navarro, Frenk, and White (NFW) profile~\cite{nfw}.
The mass enclosed within a radius within which the average density is
$200$ times as large as the critical density, $M_{200}$, will be set to
$1.4\times 10^{15} M_\odot$ and the concentration $c$ to $4.64$. I will 
assume we have ellipticities for thirty background galaxies per square arcminute,
all at redshift one. These parameters will allow us to compare with the results
of Ref.~\cite{hoekstra}.

{\parindent0pt {\it Effects of Large scale structure.}}
To get a feel for some numbers, the shear due to this cluster has magnitude
equal to $0.02$ at an angular distance $10'$ away 
from the center.
In a one square arcminute pixel with thirty background galaxies,
the rms noise due to the galaxy uncertainties is $0.25/\sqrt{30}=0.046$, significantly
larger than the signal. The rms noise due to large scale structure is 
of order $0.02$ for a standard $\Lambda$CDM model. Adding the two sources
of noise in quadrature leads to a ten percent increase in the noise due
to large scale structure.
Hoekstra~\cite{hoekstra} however showed that the situation is not that simple.
Since the signal to noise in each pixel is small, it is necessary to average over
many pixels with the same signal, in this radially symmetric case, over azimuthal
angle at fixed radius. 

If we average over all pixels in an annulus of width $1'$ at radius of $10'$, then
the shape noise gets reduced by a factor of $N_p^{-1/2}$ where $N_p$ is the number of pixels,
approximately equal to $20\pi$. The shape noise contribution to this angular average then
should be of order $0.046/(20\pi)^{1/2} \simeq 0.006$, significantly smaller than the signal. If the noise due
to large scale structure was completely independent from one pixel to the next, then it
too would be reduced by the same factor, and  would have an rms amplitude of $0.0025$. Figure~\Rf{sn} shows that shape noise does behave as expected, but the noise
due to large scale structure is larger than anticipated by more than a factor of two. Apparently,
large scale structure produces a signal (noise) which is correlated over many pixels. The number of
independent pixels then is much smaller than $20\pi$, and the contamination of the cluster signal
from large scale structure is much more severe than one might naively estimate. As shown in Figure~\Rf{sn},
the situation rapidly gets worse at larger angles, so Hoekstra argued that measurements beyond $15'$
were useless for determining the cluster mass profile: the noise from large scale structure 
overwhlems the useful information in such measurements.

\Sfig{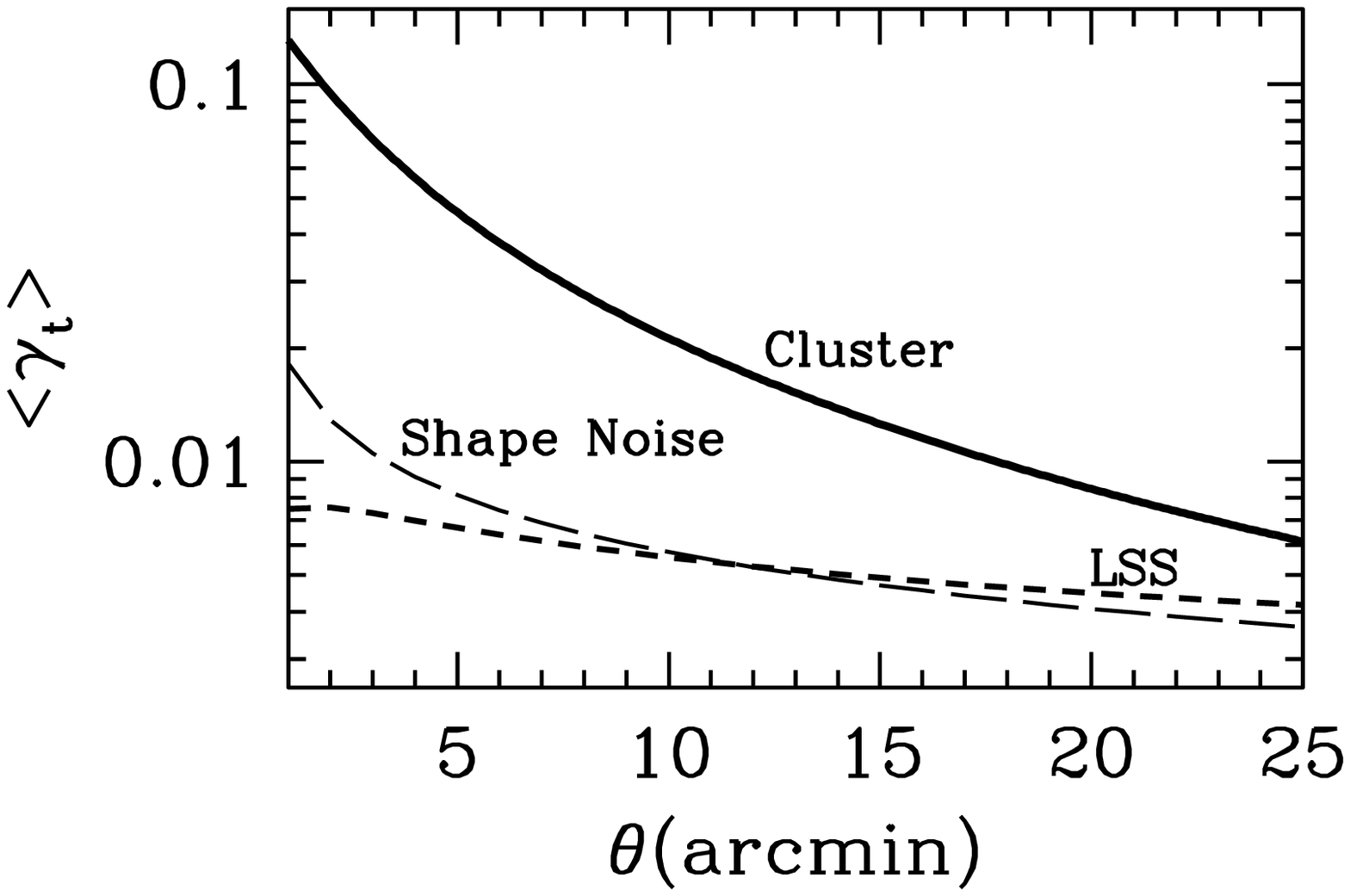}{Azimuthally averaged tangential shear~\cite{gamt}. Also shown is the rms noise due to uncertainty in 
background galaxy shapes and
due to large scale structure.}

The above hand-waving argument suggests a possible solution. Since the large scale structure signal
is correlated over $\sim 4'$, why not make use of this correlation in the radial direction as well?
That is, when averaging over the azimuthal shear, one implicitly discards all information about the
combined radial/azimuthal structure of the shear. A circular blob of shear with diameter $4'$ due to large 
scale structure might be easily detected before the angular averaging and then subtracted off. If we first
average over angles, we lose much of our power to identify and account for the shear due to large scale
structure.  

\begin{figure}[ht]
\centerline{\vbox{ 
    \includegraphics[width=0.7\columnwidth]{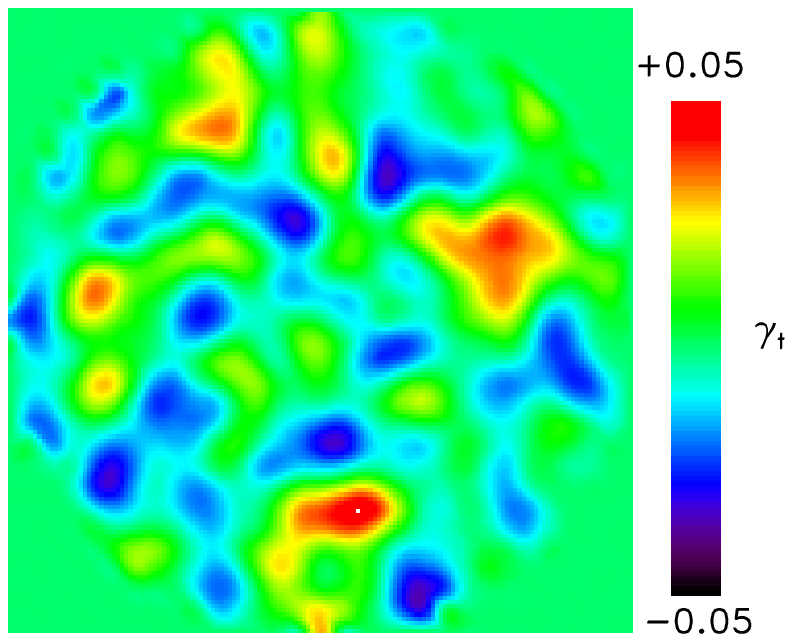}
    \includegraphics[width=0.7\columnwidth]{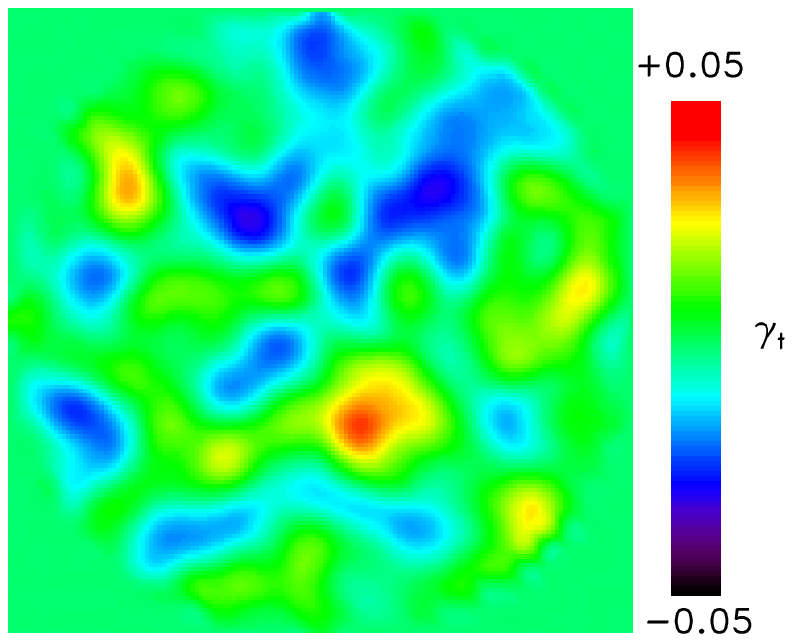}
    \includegraphics[width=0.7\columnwidth]{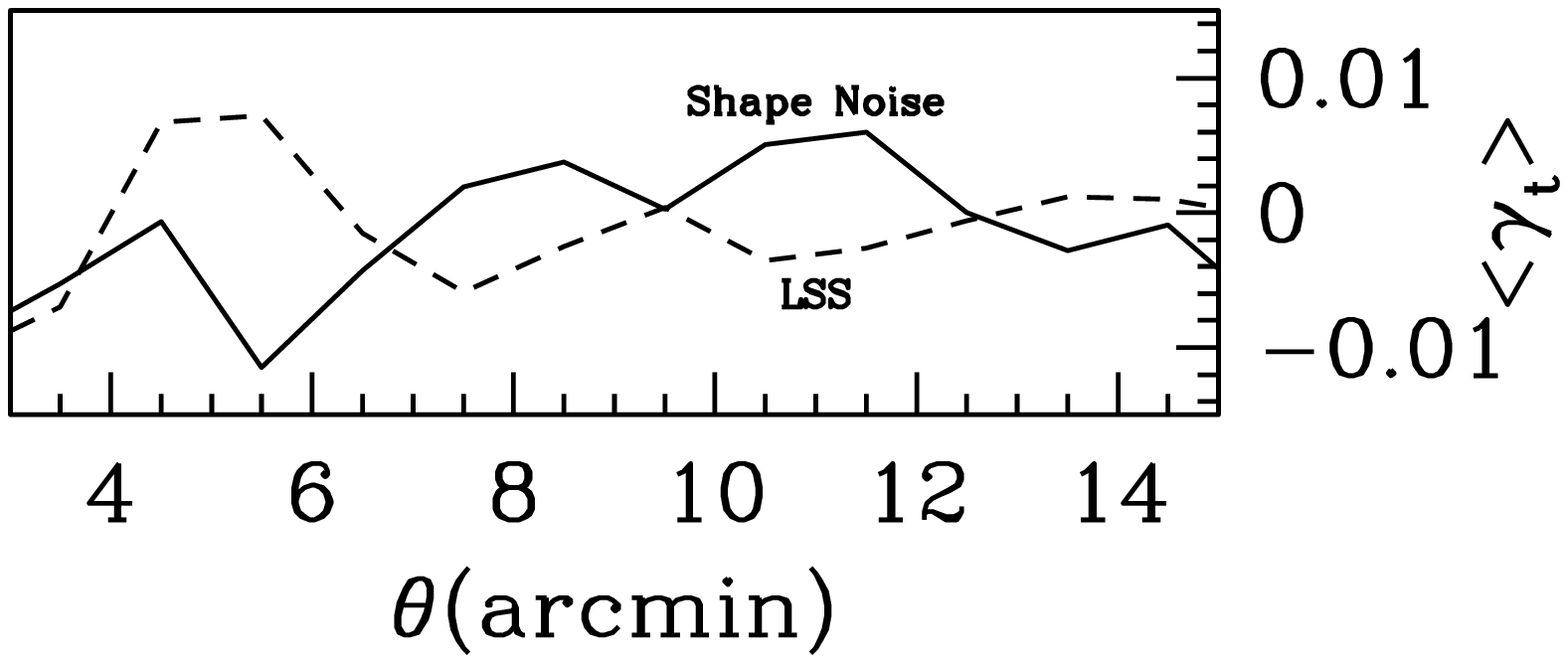}
    }
  }
\caption{{\it Top panel:} Tangential shear $\gamma_t$ due to a realization of
shape noise; {\it Middle panel:} $\gamma_t$ from large scale structure 
in a single realization. The noise emanating from large
scale structure is much more coherent than that due to the uncertainty
in galaxy shapes. Note especially the swath of negative $\gamma_t$ in the bottom
panel extending from lower left to upper right. {\it Bottom Panel:} The same
two realizations smoothed over azimuthal angle. There is no longer any clear
distinction between the two sources of noise.}
\label{fig:gamma}
\end{figure}

This is illustrated in Fig.~\Rf{gamma}, where a realization of
each source of noise is shown in the top two panels. The noise
from large scale structure is manifestly more coherent, extending over
larger scales than is the shape noise. Just as clearly, this distinction
disappears (bottom panel) if one averages over azimuthal angle.

In order to retain the information necessary to distinguish large scale structure shear from
cluster shear, we need to start from the full shear field,
$\gamma_i(\vec\theta_a)$, where $i=1,2$
labels the two components of the symmetric, traceless shear tensor. Assuming that
the shear induced by large scale structure is Gaussian, the likelihood of obtaining the data is
\be
{\cal L} = {[2\pi]^{-N_p} \over \det(N)^{1/2}} \exp\left\{
-\frac{1}{2} \big( \gamma_\alpha - \gamma_\alpha^{\rm cl} \big)
[N^{-1}]_{\alpha\beta} \big( \gamma_\beta - \gamma_\beta^{\rm cl} \big)
\right\}\eql{like}
\ee
where the index $\alpha$ includes both pixel position ($a$) and shear component ($i$)
so ranges from 1 to $2N_p$, with $N_p$ now the total number of pixels for which
we have ellipticities. The noise matrix $N$ is the sum 
of the covariance due to shape noise and to large scale structure,
\be
N \equiv C^{\rm shape} + C^{\rm lss}
.\ee
From this starting point, there are a number of directions one can take. 
Here I pursue two: (i) using the likelihood function and the Fisher matrix derived from
it to project uncertainties on the two parameters characterizing the NFW profile
and (ii) compressing the information in the likelihood function into a mass estimator
which is optimal in the sense that it retains all the relevant information. In both cases,
we can compare the results with what one would obtain without accounting for large scale
structure.

{\parindent0pt {\it Parametric Fitting.}}
\begin{figure}[ht]
\centerline{\vbox{
\includegraphics[width=.8\columnwidth]{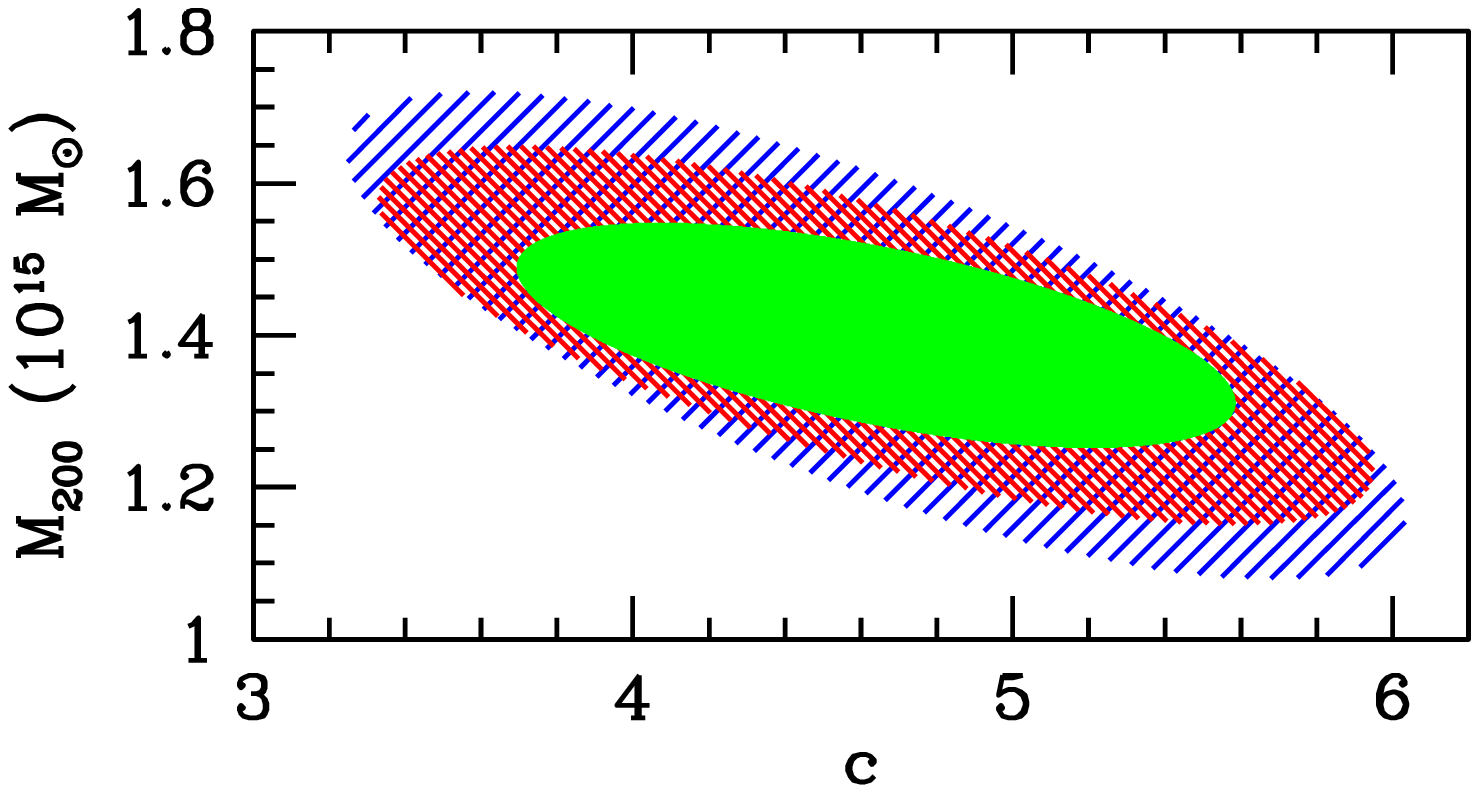}
\includegraphics[width=.8\columnwidth]{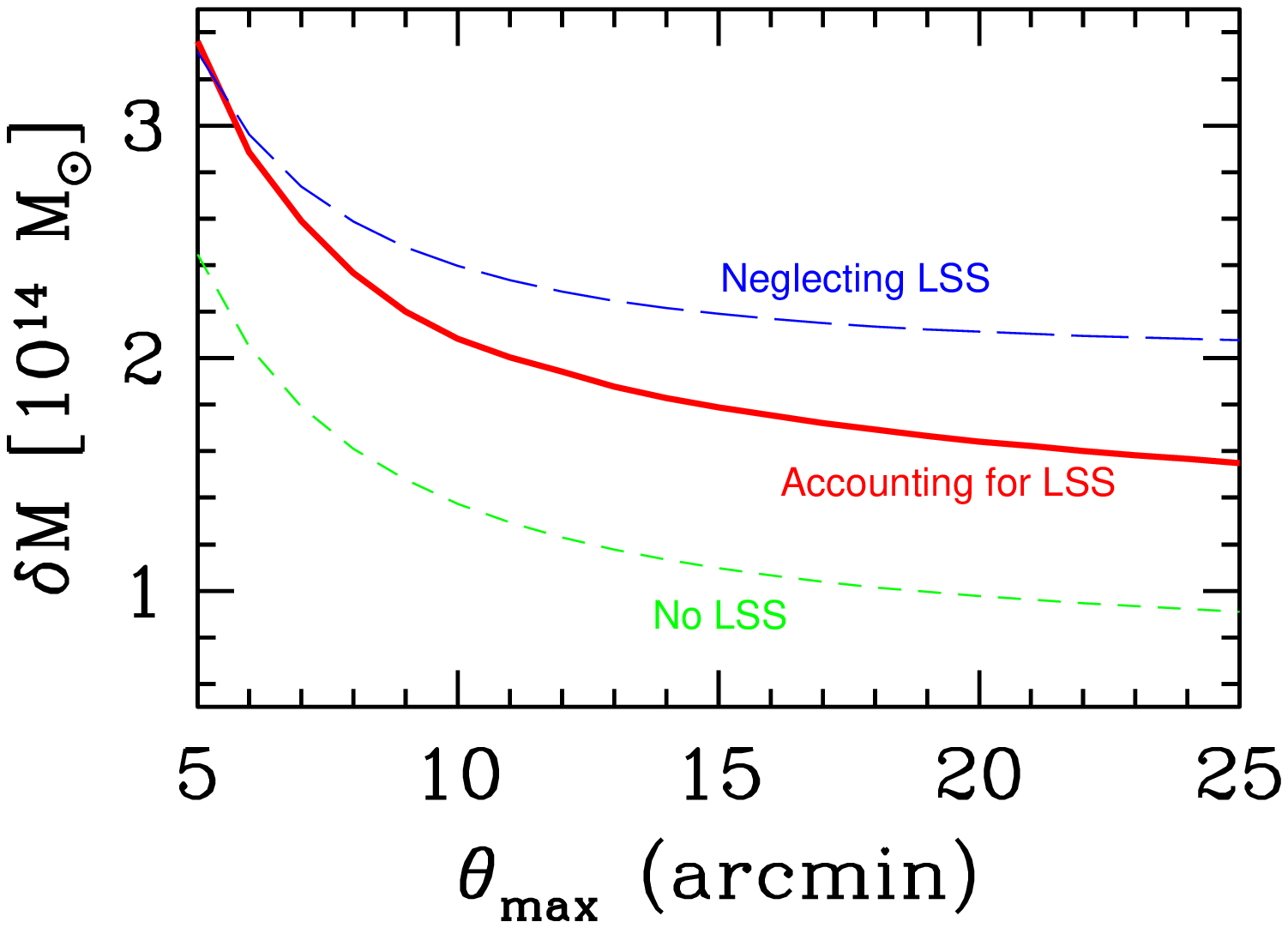}
}}
\caption{{\it Top panel:} Projected 1-sigma errors on $M_{200}$ and concentration
from ellipticities within $20'$ of the cluster
assuming $30$ background galaxies per square arcminute, all at redshift 
1. Inner ellipse assumes there is no contribution from large scale structure.
Outer ellipse shows the increase in errors once large scale structure is
included. Middle ellipse also includes large scale structure, but here the
parameters are determined directly from the likelihood function, instead of from
$\langle\gamma_t\rangle$. {\it Bottom panel:} Errors on the mass of the cluster
as a function of the maximum angle out to which ellipticities of
background galaxies are measured.}
\label{fig:ell20}
\end{figure}
Let's assume that the cluster we are studying has an NFW profile and ask how well we can
determine the parameters characterizing this profile. The Fisher matrix,
\be
F_{ij} \equiv - {\partial^2 {\cal L} \over \partial p_i\partial p_j}
= {\partial\gamma^{\rm cl}_\alpha \over\partial p_i} [N^{-1}]_{\alpha\beta} 
{\partial\gamma^{\rm cl}_\beta\over \partial p_j}
,\eql{fish}\ee
quantifies the errors on these parameters $p_1 (M_{200})$ and $p_2 (c)$
from a lensing survey. We will consider the errors on $M_{200}$ and $c$ in three
cases: (i) no large scale structure, $N=C^{\rm shape}$; (ii) including and accounting
for large scale structure $N=C^{\rm shape} + C^{\rm lss}$; and (iii)  large scale
structure present, but parameters determined from the azimuthally averages tangential
shear\cite{gamt}, $\langle\gamma_t\rangle$. 
In this last case, the data points $\gamma_\alpha$ in \ec{fish} are replaced by $\langle\gamma_t\rangle$
in different radial bins and the noise matrix $N_{\langle\gamma_t\rangle}$ consists of the shape noise 
per pixel reduced by the number of pixels in the azimuthal average and the azimuthally averaged large
scale structure covariance matrix.
Incidentally, if no large scale structure is present (i), then using all the pixels is
not necessary: $\langle\gamma_t\rangle$ maintains all the relevant information.

The top panel of Figure~\Rf{ell20} shows the projected errors from a survey which measures 
background galaxies within $20'$ of the cluster center. The difference between the inner and outer ellipse
reinforces the point emphasized by Hoekstra~\cite{hoekstra}, that the noise due to large
scale structure degrades the parameter determination by a factor of order two if not
properly accounted for. The middle
ellipse shows though that the situation is not quite this dire. If one works from
the likelihood function directly and does not azimuthally average, the effects
of large scale structure can be minimized.

The bottom panel of Fig.~\Rf{ell20} shows the error on $M_{200}$ 
(after marginalizing over the concentration)
in these three cases as a function of the maximum distance out to which 
data is available.
Here we can see that the optimal estimator (i.e. not azimuthally averaging) 
reduces the deleterious effects of large scale structure by roughly $50\%
$
at least on large scales. This realization has practical implications: is it
worthwhile taking data far from the cluster center or does the noise due
to large scale structure make such data irrelevant when it comes to determining
the cluster mass? Figure~\Rf{ell20} shows that going out to $25'$
leads to a fifteen percent smaller error on $M_{200}$ as compared with
$\theta_{\rm max}=15'$. While this is not quite as large as the $18\%
$ gain if there was no large scale structure, it is much larger
than five percent gain if large scale structure is not accounted for.

\begin{figure}[ht]
\centerline{\vbox{ \vspace{0.0in}
    \includegraphics[width=0.7\columnwidth]{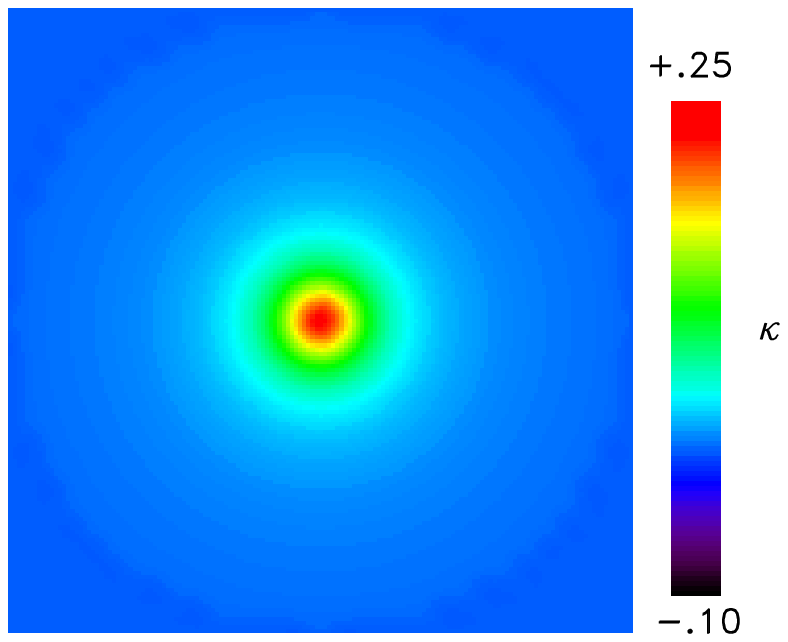}
    \includegraphics[width=0.7\columnwidth]{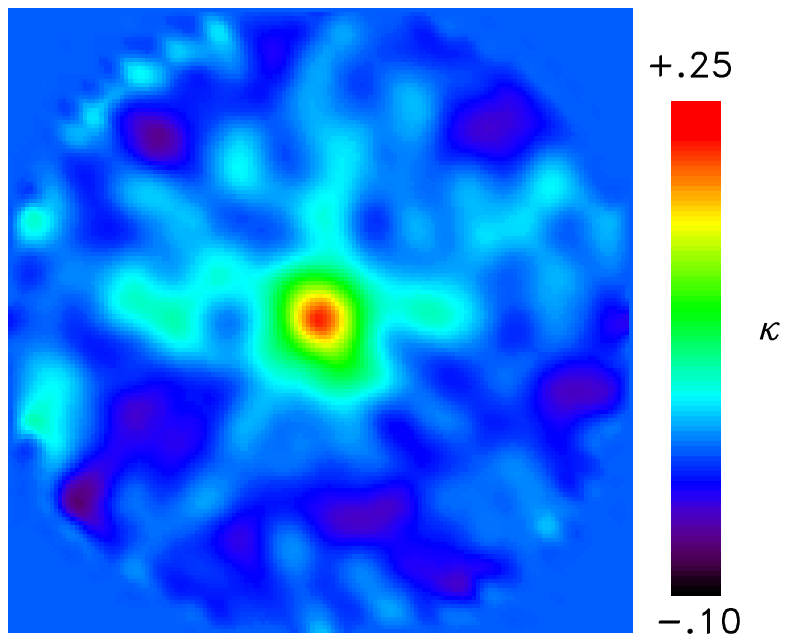}
    \includegraphics[width=0.7\columnwidth]{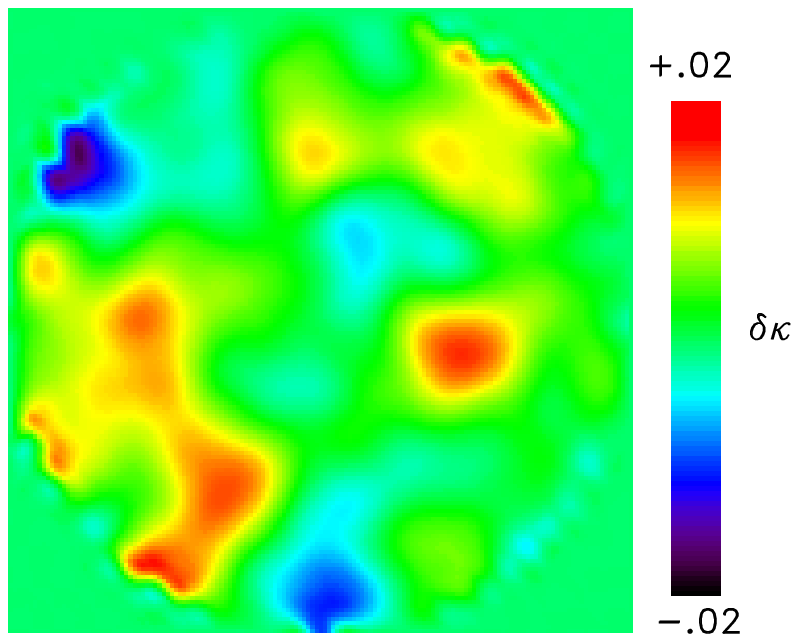}
    }
  }
\caption{Convergence of the cluster smoothed with a $1'$ Gaussian.
{\it Top panel:} Input profile ;
{\it Middle panel:} profile
recovered from simulated ellipticities accounting for large scale
structure.  {\it Bottom panel:} The difference between the LSS estimator and the estimator which 
neglects LSS. }
\label{fig:maps}
\end{figure}
{\parindent0pt {\it Mapmaking}.}
One can also try to obtain information about the mass distribution in a cluster
by inverting directly without assuming any particular profile. Since Kaiser
and Squires first introduced this idea~\cite{squireskaiser}, many groups have worked
to develop new techniques accounting for real world complications~\cite{schneider}. 
Here I simply want
to discuss a way to improve an estimator by accounting for large scale structure. So, I will
focus on one particular estimator~\cite{hukeet}, the so-called optimal estimator, used recently to make
maps in CMB experiments~\cite{mc}. Although the maps made from this estimator are not
particularly pretty, they do retain all the information stored
in the likelihood function. Thus, they are very useful for quantitative analysis and they can
be massaged in a number of ways (which I will not discuss here) to produce more realistic
pictures. The main point is to see how much we can learn in the face of noise due
to large scale structure.

The shear due to a cluster is linearly related to the convergance 
\be
\gamma^{\rm cl}_\alpha = K_{\alpha a} \kappa_a
\ee
where the index $a$ 
ranges over all pixels for which we are fitting the surface density. This
presumably will be of order $N_p$ (but it does not have to be exactly equal to
it, for we may choose to estimate the density in a coarser grid than the
measured ellipticities). The kernel 
is
\be
K_{\alpha a} = -{A\over\pi\vert\theta_\alpha-\theta_a\vert^2} 
\cases{ \cos(2\phi) & $i=1$\cr
-\sin(2\phi) & $i=2$}
\ee
where $\phi$ is the angle between the $x$-axis
and the vector $\theta_\alpha-\theta_a$, and $A$ is the area of a pixel.

The measured shear is a combination of this signal,
shape noise, and projections from large scale structure:$
\gamma_\alpha = \gamma^{\rm cl}_\alpha + \gamma^{\rm shape}_\alpha + \gamma^{\rm lss}_\alpha
.$
The latter two contributions have mean zero and a total covariance matrix $N$.
Then the minimum variance estimator for the convergence is
\be
\hat\kappa = C_N K^t N^{-1} \gamma
\eql{mvm}\ee
with mean equal to $\kappa^{\rm cl}$ and covariance matrix
\be
C_N \equiv \left( K^t N^{-1} K\right)^{-1} 
.\ee

Figure~\rf{maps} shows the convergance of the cluster, 
along with the reconstruction accounting for large
scale structure. Each of these was obtained from
a pixelized set of ellipticities out to a radius of $15'$ which included 
the signal due to the cluster, shape noise, and large scale structure. 
Fig.~\Rf{gamma} shows the two noise sources. The map in the
middle of Fig.~\Rf{maps} was obtained with the minimum variance estimate of \ec{mvm}.

The bottom panel of Fig.~\Rf{maps} shows the difference between a map which
accounts for large scale structure and one that does not (i.e., one in which
$N$ was set to $C^{\rm shape}$). Note especially
that the minimum variance estimator obtains greater densities along a swath extending
from the lower left to upper right. A comparison with the top panel shows that
the minimum variance estimator more accurately reproduces the density. 
It does this by properly accounting for large scale structure. The estimator
which neglects LSS treats the negative $\gamma_t$ swath from lower left to upper
right in Fig.~\Rf{gamma} as produced by the cluster. Thus, the total $\gamma_t$ it ascribes
to the cluster is smaller than it should be. The result is an underprediction of the
density along this swath. The minimum variance estimator avoids this pitfall.

How much does the variance go up when one uses the estimator
which neglects large scale structure?
\Sfig{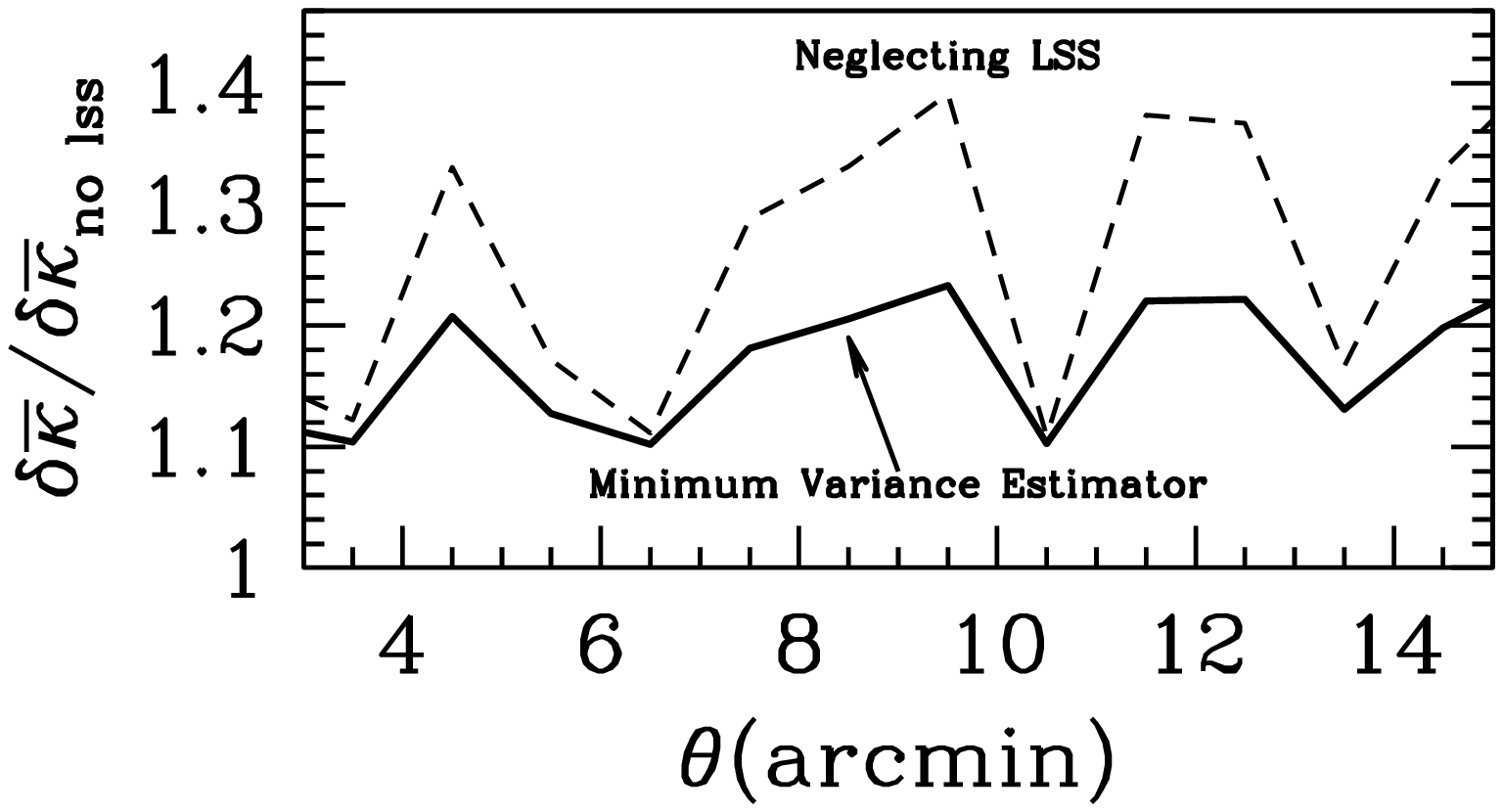}{Uncertainty in the mass enclosed in an annulus of radius $\theta$
and width $1'$ from estimators compared with the uncertainty in the absence of
large scale structure. Dashed curve uses the equivalent of \ec{mvm}, without
accounting for large scale structure.}
Fig.~\Rf{annulus} shows the errors on the mass within an annulus of radius
$\theta$ compared with the errors if there was no large scale structure. As in
the parametric estimation, we see that the minimum variance estimator more
accurately determines the cluster mass. While the errors are larger than if
there had been no large scale structure, they are significantly smaller than one
gets when neglecting large scale structure. 

This work is supported by the DOE, by NASA grant NAG5-10842, 
and by NSF Grant PHY-0079251. I am grateful to H.~Hoekstra,
C.~Keeton, E.~Rozo, and M.~White for helpful comments.

\newcommand\spr[3]{{\it Physics Reports} {\bf #1}, #2 (#3)}
\newcommand\sapj[3]{ {\it Astrophys. J.} {\bf #1}, #2 (#3) }
\newcommand\saa[3]{ {\it Astron. \& Astrophys.} {\bf #1}, #2 (#3) }
\newcommand\sprd[3]{ {\it Phys. Rev. D} {\bf #1}, #2 (#3) }
\newcommand\sprl[3]{ {\it Phys. Rev. Letters} {\bf #1}, #2 (#3) }
\newcommand\np[3]{ {\it Nucl.~Phys. B} {\bf #1}, #2 (#3) }
\newcommand\smnras[3]{{\it Monthly Notices of Royal
	Astronomical Society} {\bf #1}, #2 (#3)}
\newcommand\splb[3]{{\it Physics Letters} {\bf B#1}, #2 (#3)}


\begin{thebibliography}{99}

\bibitem{cluster} J.~P.~Henry, \sapj{489}{L1}{1997}; P.~T.~P.~Viana,
and A.~R.~Liddle, \smnras{303}{535}{1999}; T.~H.~Reiprich and
H.~Bohringer, \sapj{567}{716}{2002}; V.~R.~Eke, S.~Cole, and C.~S.~Frenk,
\smnras{282}{263}{1996}; N.~A.~Bahcall and X.~Fan, \sapj{504}{1}{1998}

\bibitem{wang} L.~Wang and P.~J.~Steinhardt, \sapj{508}{483}{1998}

\bibitem{haiman} Z.~Haiman, J.~J.~Mohr, and G.~P.~Holder, \sapj{553}{545}{2001}

\bibitem{scatter} In the case of optical richness, see, e.g., the theoretical
investigations in M.~P.~van Haarlem, C.~S.~Frenk, and S.~D.~M.~White, \smnras{287}{817}{1997}; 
K.~Reblinsky and M.~Bartelmann, \saa{345}{1}{1999}; M.~White and C.~S.~Kochanek,
\sapj{574}{24}{2002}

\bibitem{bartel} For a review see M.~Bartelmann and P.~Schneider,
\spr{340}{291}{2001}. An incomplete list of weak lensing mass determinations 
is:
J.~A.~Tyson, R.~A.~Wenk, and F.~Valdes, \sapj{349}{L1}{1990}; Th.~Erben et al.,
\saa{355}{23}{2000}; D.~Clowe et al., \sapj{539}{540}{2000};
P.~J.~Marshall et al.,  \smnras{335}{1037}{2002}.

\bibitem{los} C.~A.~Metzler, M.~White, and C.~Loken, \sapj{547}{560}{2001};
M.~White, L.~van Waerbeke, and J.~Mackey, \sapj{575}{640}{2002}

\bibitem{hoekstra} H.~Hoekstra, \saa{370}{743}{2001}, \smnras{339}{1155}{2003}

\bibitem{nfw} J.~F.~Navarro, C.~S.~Frenk, and S.~D.~M.~White, \smnras{275}{720}{1995},
\sapj{490}{493}{1997}


\bibitem{gamt} $\gamma_t\equiv -\gamma_1\cos(2\phi)-\gamma_2\sin(2\phi)$ where $\phi$ is the 
azumthal angle $\theta$ makes with a fixed $x$-axis. 

\bibitem{squireskaiser} N.~Kaiser and G.~Squires, \sapj{404}{441}{1993};
G.~Squires and N.~Kaiser, \sapj{473}{65}{1996}.

\bibitem{schneider} See, e.g., P.~Schneider, {\tt astro-ph/0306465}

\bibitem{hukeet} W.~Hu and C.~R.~Keeton, \sprd{66}{063506}{2002}

\bibitem{mc} See, e.g., S.~Dodelson, {\it Modern Cosmology} (Academic Press, San Diego, 2003)

\end{thebibliography}
\end{document}